\pdfoutput=1
\documentclass[showpacs,prb,twocolumn]{revtex4}
\usepackage{amsmath}
\usepackage{amssymb}
\usepackage{color}
\usepackage{graphics}
\usepackage{graphicx}
\usepackage{dcolumn}

\begin{document}

\title {Surface plasmon enhanced absorption and suppressed transmission in periodic arrays of graphene ribbons}

\author{ A.~Yu.~Nikitin$^{1,2}$}
\email{alexeynik@rambler.ru}
\author{ F. Guinea$^3$}
\author{ F.~J.~Garcia-Vidal$^4$}
\author{ L.~Martin-Moreno$^1$}
\email{lmm@unizar.es}
 \affiliation{$^1$ Instituto de Ciencia de Materiales de Arag\'{o}n and Departamento de F\'{i}sica de la Materia Condensada,
CSIC-Universidad de Zaragoza, E-50009, Zaragoza, Spain \\
$^2$A.Ya. Usikov Institute for Radiophysics and Electronics, Ukrainian Academy of
Sciences, 12 Acad. Proskura Street, 61085 Kharkov, Ukraine\\
$^3$ Instituto de Ciencia de Materiales de Madrid, CSIC, Cantoblanco, E-28049 Madrid, Spain\\
$^4$ Departamento de F\'{i}sica Te\'{o}rica de la Materia Condensada, Universidad Aut\'{o}noma de Madrid, E-28049, Madrid, Spain}

\begin{abstract}

Resonance diffraction in the periodic array of graphene micro-ribbons is theoretically studied following a recent experiment [L. Ju et al, Nature Nanotech.  \textbf{6}, 630 (2011)]. Systematic studies over a wide range of parameters are presented.  It is shown that a much richer resonant picture would be observable for higher relaxation times of charge carriers: more resonances appear and transmission can be totally suppressed. The comparison with the absorption cross-section of a single ribbon shows that the resonant features of the periodic array are associated with leaky plasmonic modes. The longest-wavelength resonance provides the highest visibility of the transmission dip and has the strongest spectral shift and broadening with respect to the single-ribbon resonance, due to collective effects.

\end{abstract}

\pacs{42.25.Bs, 41.20.Jb, 42.79.Ag, 78.66.Bz} \maketitle

The ability of graphene to support electromagnetic waves coupled to charge carriers [graphene surface plasmons (GSPs)] is very interesting from the point of view of many physical phenomena related to surface plasmons (SPs)\cite{BarnesNature03,Maierbook}. An additional interest is related to graphene's flexibility, sensitivity to external exposure and two-dimensionality (2D) that have a variety of possible applications.\cite{Review09,graphphot_natphot10} GSPs have been intensively studied theoretically,\cite{Shung86,Campagnoli89,Vafek06,Hansonw08,IRSPP09,KoppensNanoLett11,NikitinPRB11}
in graphene sheets,  and also in graphene ribbons\cite{BreyPRB07,SilvestrovPRB08,ribbonsPRB10,Engheta,NikitinPRBR11,AndersenArXive11}, p-n junctions\cite{p-n_junction} and edges\cite{Engheta,NikitinPRBR11,edgePRB11} and recently have been observed experimentally.\cite{PlasmonicsNature11,FeiNanolett11}

In metal films, the excitation of the SP modes had been experimentally and theoretically studied for periodic ultrathin structures ($\lesssim10$ nm-thick), both for arrays of slabs\cite{SpevakPRB09,XiaoAPL10,AguannoPRB11,Plasmonics11} and arrays of holes and disks\cite{RodrigoOL09,BraunPRL09,XiaoOL11}. It has been shown that these systems present transmission peaks with high visibility (including total suppression of reflection) and absorption resonances. The natural continuation of this research was to check if this property could still hold for the 2D limit i.e., for a layer of one-atom thickness. Recently, experiments have shown that GSP resonances in a periodic array of graphene ribbons (PAGR) have remarkably large oscillator strengths, resulting in prominent room-temperature optical absorption peaks.\cite{PlasmonicsNature11}.

\begin{figure}[thb!]
\includegraphics[width=7cm]{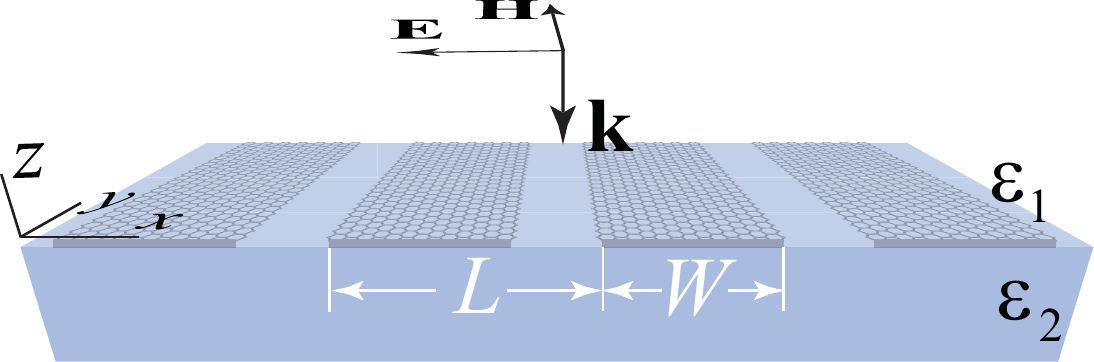}\\
\caption{(Color online)  The geometry of the studied system: a periodic array of graphene micro-ribbons of width $W$ and period $L$, with a normally-incident electromagnetic wave having the magnetic field along the ribbons. The array is placed between two dielectric half-spaces with dielectric constants $\varepsilon_1$ and $\varepsilon_2$.}\label{geom}
\end{figure}

In this paper we present a theoretical study of the electromagnetic response of PAGRs, including absorption, transmission and reflection coefficients. We consider both the parameters corresponding to the experiment and their variation over a wide range. Specifically, we focus on the dependencies upon the relaxation times of charge carriers $\tau$ and the width-to-period ratio (which in the experiment was fixed to be $1/2$). We look for the configurations in which GSP-induced absorption is enhanced and where other GSP-assisted effects are much more pronounced. Our analysis can thus be used for further efficient observation of GSPs and their use for applications e.g., ultra-thin voltage-controllable THz absorbers.

Figure~\ref{geom} schematically represents the periodic array  of graphene ribbons under study. The PAGR is located at $z=0$ and is illuminated by a normal-incident monochromatic plane wave
(having vacuum wavelength $\lambda$), with electric field pointing along the $x-$direction. The period of the PAGR is $L$, the width of the ribbon is $W$ and the dielectric permittivities of the superstrate and substrate are $\varepsilon_1$ and $\varepsilon_2$, respectively. The graphene ribbons are modeled using a 2D conductivity $\sigma$, computed within the random-phase approximation.\cite{Wunsch06,Hwang07,Falkovsky08} Room temperature, $T=300$K is considered throughout the paper.

\begin{figure}[tbh!]
\includegraphics[width=8.3cm]{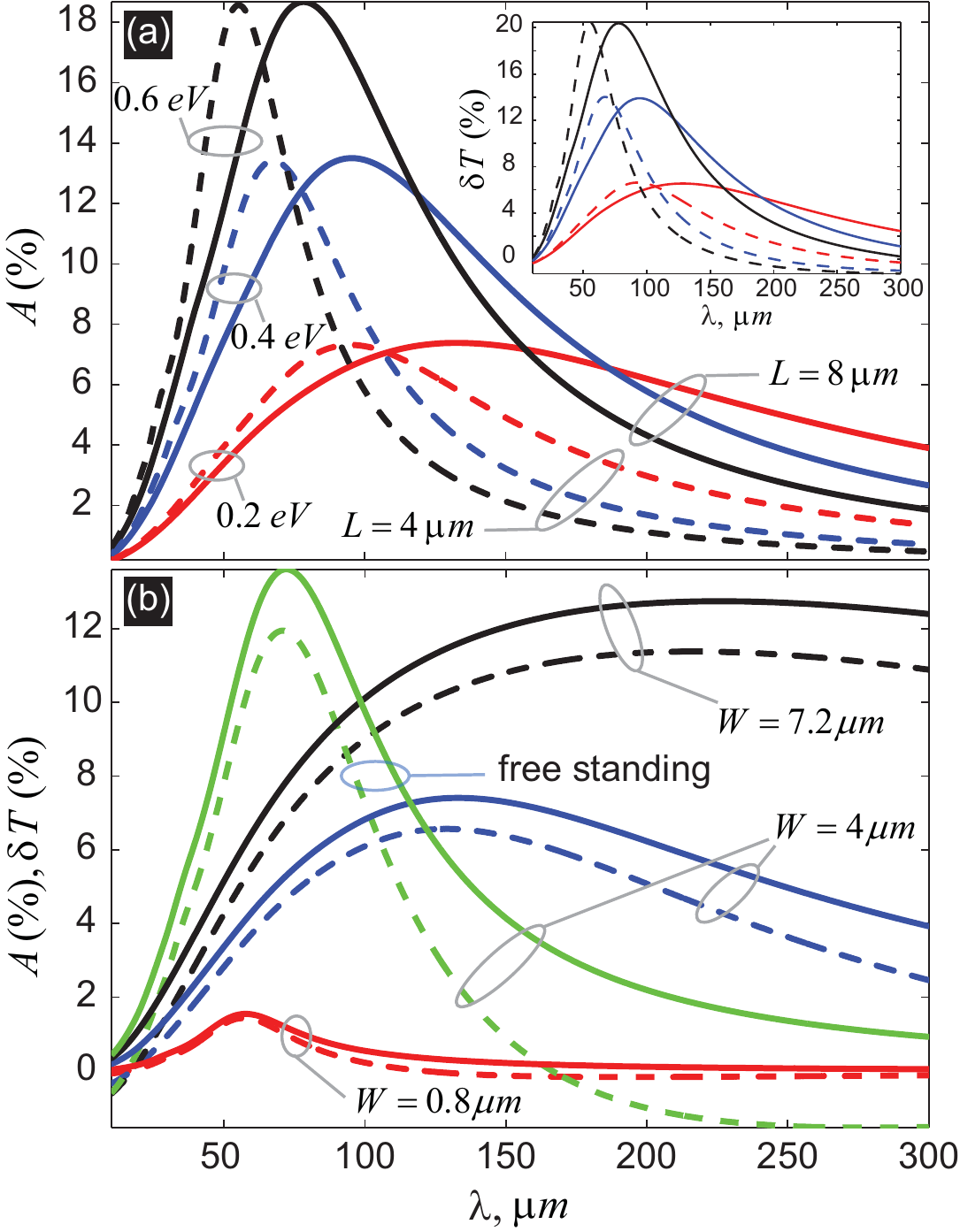}\\
\caption{(Color online) Absorption and gate-induced change of transmission spectra for PAGR with the dielectric cladding and relaxation time corresponding to the experiment\cite{PlasmonicsNature11}: $\varepsilon_1=3$, $\varepsilon_2=4$, $\tau=0.25$ ps. (a) shows the absorption spectra $A$ for different values of chemical potential $\mu$. The continuous curves are for the period $L = 8 \mu $m, while the discontinuous ones are for $L = 4 \mu $m. For both cases $W=L/2$. In the inset the relative change of transmission with respect to the sample at charge neutral point $\delta T = -(T-T_{CNP})/T_{CNP}$ is shown for the same $L$ and $W$ as in the main figure. In (b) the spectra for $A$ (continuous curves) and $\delta T$ (discontinuous curves) are shown for different widths of the ribbons $W$ for $L = 8 \mu $m, $\mu =0.2$eV. The case of $W = 4 \mu $m is also shown for the free-standing graphene ($\varepsilon_1=\varepsilon_2=1$).}\label{exper}
\end{figure}

Due to diffraction, the PAGR generates an infinite discrete set of plane waves $n\in \mathbb{Z}$ with $x$-components of the wavevectors $k_{nx}=nG$, $G=2\pi/L$ being the shortest vector of the reciprocal lattice. The fields in the dielectric half-spaces can then been presented in the standard form of Fourier-Floquet expansion. Matching the fields at the interface $z=0$ results in an infinite set of linear equations for the amplitudes of diffracted waves. The direct calculation of the diffraction amplitudes in the truncated linear system is simple to implement and, additionally, provides qualitative information on spectra of GSPs. However, the convergency of this procedure with respect to higher harmonic considered, $N$, is poor for the chosen polarization. Therefore, for each geometry and for the lowest scattering time $\tau$ considered, the modal expansion calculations have been checked by the finite elements method (FEM) realized in \textsc{comsol}. Once the value of $N$ needed to achieve convergency is found, the faster modal expansion method can be used to study the dependency with $\tau$ of the scattering coefficients.

Let us start our analysis of the electromagnetic properties of PAGR by the geometry considered in the experiment described in Ref.~\onlinecite{PlasmonicsNature11}. For this, we take $\varepsilon_1=3$ (ion gel) and $\varepsilon_2=4$ ($\mathrm{SiO}_2$), ignoring thus the effect of a finite thickness of the
$\mathrm{SiO}_2$ layer and possible related Fabry-Perot type resonances. The scattering rate is assumed to be $4$THz ($\tau=0.25$ps). As in the experiment, the transmission coefficient is compared with the one at the ``charge neutral point'' $T_{CNP}$, where the chemical potential is very small  (we take $\mu =10^{-2}$eV).

\begin{figure*}[tbh!]
  \includegraphics[width=16cm]{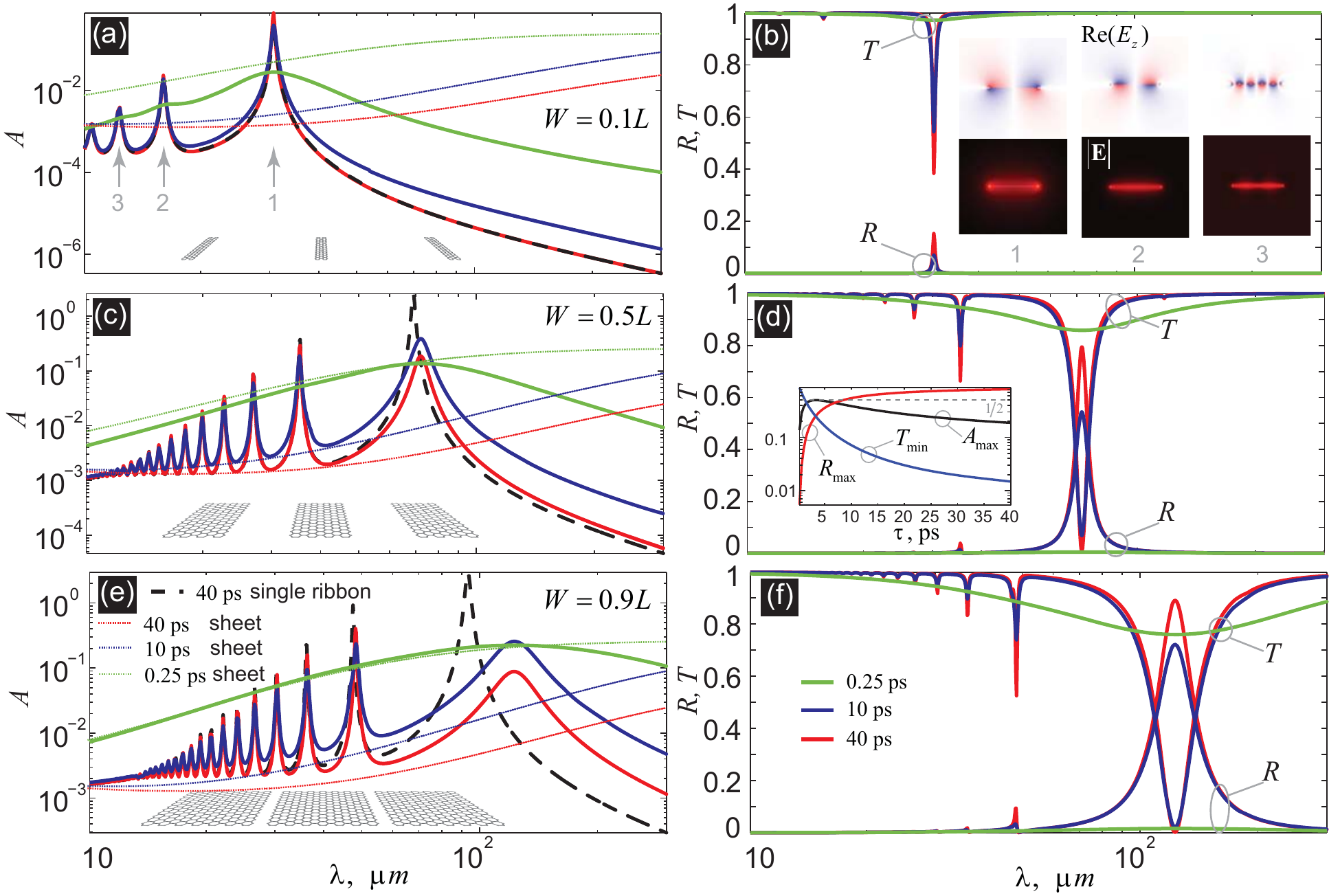}\\
\caption{(Color online) Absorption $A$ [in panels (a),(c),(e)], reflection $R$ and transmission $T$ [in panels (b),(d),(f)] spectra for free-standing periodic array of graphene ribbons with different values of ribbon width $W$ and relaxation times $\tau$. Panels (a)-(b), (c)-(d), (e)-(f) correspond to $W=0.8$, $4$ and $7.2$$\mu$m, respectively; in all cases the period is $L=8$$\mu$m. The inset to (b) represents the colorplots for the electric field modulus $|\mathbf{E}|$ and the real part of the $y$-component of the electric field $E_y$ in the vicinity of the ribbon. The colorplots are marked with respect to the numeration of the peaks in panel (a). The inset to (d) shows the dependencies of the maximum values of $A$ and $R$ and the minimum value of $T$ as a function of $\tau$, for longest-wavelength resonance shown in (c)-(d). The horizontal discontinuous line in the inset sets the maximal possible value of $A$. The dotted lines in panels (a),(c),(e) represent the absorption by a continuous graphene sheet.}\label{spectra}
\end{figure*}

First of all, following the experimental study, we consider the variation of the spectra with the change of the chemical potential $\mu$ and the period $L$, for a fixed ratio, $W/L=1/2$. Fig.~\ref{exper} (a) shows the transmission coefficient change $\delta T = -(T-T_{CNP})/T_{CNP}$ and absorption coefficient $A$ as a function of the wavelength. For each value of $L$ and $\mu$ there is a resonant maximum in both $A$ and $\delta T$ spectra. As will be seen below, the resonance is related to the excitation of the longest-wavelength GSP in each ribbon. The maximum resonant absorption increases with the increase of doping, due to both the resonance shift to a less absorptive frequency region and to the higher number of charge carriers that get involved in the plasmonic oscillation. In accordance with the experiment, this resonance blue-shifts when either $\mu$ increases or $L$ decreases.  This behavior can be explained from the condition for GSP resonance in the ribbon, which approximately satisfies\cite{NikitinPRBR11}  $W\sim n\lambda_{GSP}/2$. Here $\lambda_{GSP}=2\pi/\mathrm{Re}(k_{GSP})$ is the GSP wavelength, and $n$ measures the number of half-wavelengths that fit within the ribbon width for a certain mode. In the considered frequency range, the intra-band Drude-like term dominates in the conductance, so $\mathrm{Re}(k_{GSP})\simeq\hbar\omega^2/(2\alpha_0\mu c)$, where $\alpha_0$ is the fine-structure constant. Substituting $\mathrm{Re}(k_{GSP})$ into the resonance condition, we have for the resonance wavelength $\lambda_{res}\sim\sqrt{2\pi c \hbar W/(n\alpha_0\mu)}\propto \sqrt{W/\mu}$.

Further insight into the absorption process can be gained from going beyond the $W/L=1/2$  ratio considered in the experiment\cite{PlasmonicsNature11}. Figure~\ref{exper}(b) shows the spectra for $A$ and $\delta T$
for different widths of the ribbons $W$, at the fixed period $L=8\mu$m and for $\mu=0.2$eV. For larger values of $W$ the resonance shifts to longer wavelengths, where graphene is more absorptive and, correspondingly, the peak broadens. Interestingly, the propagation length of GSPs increases at longer wavelengths since the increase of $\mathrm{Re}(\sigma)$ is overcompensated by the lower GSP confinement. Nevertheless and despite the presence of resonances, for the considered $\tau=0.25$ps, the maximum of $A$ grows with $W$ reaching its maximum for a continuous graphene sheet $W=L$ (this dependence is almost linear, as shown by calculations for intermediate values of $W$ not presented here). In other words, for small values of $\tau$ the effect of GSP-induced absorption is weak, so that the absorption is approximately proportional to the area covered by graphene. Actually, as will be rendered in Fig~\ref{spectra}(c), the computed absorption is higher for a continuous graphene sheet than for
a PAGR with the $W/L=1/2$  ratio considered in the experiment.

Our calculations show that the GSP-absorption effect would be greatly enhanced for higher values of $\tau$, which are currently associated to free-standing graphene sheets and their much higher electron mobilities.\cite{HwangPRL07,BolotinSSC08} In order to differentiate between the effects of changing the dielectric environment and changing the relaxation time, Fig.~\ref{exper}(b) renders $A$ and $\delta T$ for a free standing PAGR, with $W/L=1/2$ and $\tau=0.25$ps.
The resonance in the free-standing PAGR blue-shifts and is get narrower than that of the corresponding PAGR with dielectric surrounding (which is related to the shift of GSP dispersion curves), but all the discussed tendencies with the change of $\mu$ and $W$ are the same. Similarly and even though the maximum absorption in the PAGR has increased, the absorption in the spectral window considered is below the one for a continuous graphene sheet.

This situation changes at higher relaxation times. Figure~\ref{spectra} illustrates the absorption, transmission and reflection spectra for the suspended PAGR of different relaxation times $\tau=0.25,10$ and $40$ps and different widths of the ribbons $W=0.1,0.5$ and $0.9L$, for the period $L=8\mu$m.
To make a set of the resonance peaks more visible (specially those appearing at lower wavelengths) the absorption is presented in logarithmic scale. For each ribbon width the absorption coefficient is compared with the absorption cross-section ($A_{SR}$) for a single ribbon of the same width at $\tau=40$ps. For better comparison, the $A_{SR}$ is normalized so that its value coincides with $A$ at the shortest wavelength in the considered spectral interval.

Each peak on the absorption spectra corresponds to a GSP resonance in the ribbon.
Increasing $W$ increases the number of resonances that appear in the spectral window considered. These resonances correspond to the excitation of either GSP waveguide- or edge-type modes with zero value of $k$-vector in the $y$-direction. These are leaky modes, resulting from the GSPs discussed in Ref.~\onlinecite{NikitinPRBR11}, with the prolongation of the dispersion curves inside the light cone down to the value $\mathrm{Re}(k_y)=0$. The field distribution around a ribbon corresponding to the last three peaks in the absorption spectra is shown in the inset of Fig.~\ref{spectra}(a). The two highest-wavelength modes result from the degenerate edge GSPs while the rest of the resonances correspond to excited waveguide-type GSPs.\cite{NikitinPRBR11}
Notice that, in practically all cases, the absorption spectra for the array and the single ribbon are approximately equal (independently on the value of $\tau$). The only exception occurs for the resonance appearing at the
longest wavelength, and only for narrow gaps between the ribbons (gap width $\lesssim 0.2 L$), when the GSPs in neighboring ribbons hybridize.

In the symmetric dielectric environment considered, $A$ in the graphene array
can not exceed its maximal value\cite{deAbajoArXive11} $A_{M}=1/2$. But importantly,
even for small ratios $W/L$, for sufficiently large values of $\tau$ the GSP-induced absorption in PAGR can not only be higher than the absorption corresponding to lower $\tau$, but can also largely exceed the absorption in the continuous graphene sheet [see Fig.~\ref{spectra}(a),(c),(e)].

The calculations rendered in Fig.~\ref{spectra}(b),(d),(f) show that absorption peaks are complemented by peaks in reflection $R$ and dips in transmission $T$, with the longest-wavelength resonance presenting the deepest minimum in $T$.

Let us now focus on the longest-wavelength resonance at $W/L=1/2$ [see Fig.~\ref{spectra}(c) and Fig.~\ref{spectra}(d)]. The inset to Fig.~\ref{spectra}(d) renders the dependency on relaxation time of the maximum values of $A$ and $R$ together with the minimal value of $T$. As seen, the evolution of $A_{max}$ with $\tau$ is not monotonous, with $A_{max}$ reaching the optimum value at $\tau\simeq3$ps. Conversely, $T_{min}$ monotonically decreases  with $\tau$, while $R_{max}$ monotonically increases with it. Importantly, the minimal value of the transmission in the resonance  keeps its low value $T_{min}<10\%$ down to $\tau\gtrsim7$ps, and $T_{min}<2\%$ for $\tau\gtrsim30$ps. Taking into account hight values of mobilities in suspended samples,\cite{HwangPRL07,BolotinSSC08} these deep transmission minima could be observed experimentally.

To conclude, we have studied the transmission, reflection and absorption resonance THz spectra in periodic arrays of graphene ribbons. The resonance effects are related to the leaky plasmonic modes existing in individual ribbons and the modes corresponding to different ribbons are very weakly coupled to each other. The highest-wavelength resonance provides the maximal visibility of the transmission dip and reflection peak, with its resonant character surviving even for the low relaxation times present in graphene samples on a substrate. As this mode is the less confined, it is the most strongly perturbed by the periodicity of the array. The samples with higher relaxation times allow for more resonances being visible and provide very deep transmission minima. We have shown that, in ribbon arrays with sufficiently high relaxation time,  the absorption can be substantially higher than the absorption in the continuous graphene sheet.

The authors acknowledge support from the Spanish MECD under Contract No. MAT2009-06609-C02, FIS2008-00124, CONSOLIDER CSD2007-
00010,  and Consolider Project ``Nanolight.es''. A.Y.N. acknowledges the Juan de la Cierva Grant No. JCI-2008-3123.

\end{document}